\documentclass[conference,10pt]{IEEEtran}
\usepackage{epsf}
\usepackage{graphicx}
\usepackage{amsmath}
\usepackage{amssymb}
\usepackage{epsfig,latexsym,amsmath,epsf,amssymb,amsfonts}
\usepackage{algorithm, algorithmic}
\usepackage{placeins}
\usepackage{array,color}
\usepackage{cite}
\usepackage{textcomp}
\usepackage{makecell}
\usepackage{caption}
  {\end{list}}

\newcommand{\cmmnt}[1]{\ignorespaces}

\renewcommand{\theequation}{\arabic{equation}}

\begin{document}
\bstctlcite{IEEEexample:BSTcontrol}

\title{ Deep Reinforcement Learning for Distributed  \\[-.08in]
Uncoordinated Cognitive Radios\\[-.08in]
Resourec Allocation \vspace*{-6mm}}
\author{\large Ankita Tondwalkar and Andres Kwasinski \\ [-.03in]
\normalsize  Rochester Institute of Technology, Rochester, New York 14623, USA \\[-.03in]
\{at3235, axkeec\}@rit.edu \vspace*{-6mm}
}

\maketitle 

\begin{abstract}
This paper presents a novel deep reinforcement learning-based resource allocation technique for the multi-agent environment presented by a cognitive radio network that coexists through underlay dynamic spectrum access (DSA) with a primary network. The resource allocation technique presented in this work is distributed, not requiring coordination with other agents. The presented algorithm is the first deep reinforcement learning technique for which convergence to equilibrium policies can be shown in the non-stationary multi-agent environment that results from the uncoordinated dynamic interaction between radios through the shared wireless environment. Moreover, simulation results show that in a finite learning time the presented technique is able to find policies that yield performance within 3 \% of an exhaustive search solution, finding the optimal policy in nearly 70 \% of cases. Moreover, it is shown that standard single-agent deep reinforcement learning may not achieve convergence when used in a non-coordinated, coupled multi-radio scenario.
\end{abstract}
\vspace*{-2mm}

\IEEEpeerreviewmaketitle

\section{Introduction}\vspace*{-1mm}

Because of their ability to autonomously gain awareness of the wireless network environment and learn to adapt to changing conditions by applying machine learning technology, cognitive radios (CRs) are considered as the answer to realize the more efficient and effective use of the radio spectrum attainable through Dynamic Spectrum Access (DSA). Due to its model-free characteristic, reinforcement learning (RL) is the machine learning approach to resource allocation that naturally aligns with this vision for CRs. However, the application of RL in general wireless networking scenarios present the challenge that the multiple CR links are entangled through the shared spectrum, materialized by the actions of one CR (its transmissions) affecting the environment (e.g. interference) of the other CRs, and resulting in a multi-agent non-stationary environment. In this work we consider the specially challenge case of a multi-agent non-stationary environment where  CRs in a CR network (CRN) operate in an uncoordinated distributed fashion by following an underlay DSA paradigm to share the spectrum with a primary network (PN). Specifically, the goal of this work is to develop a RL mechanism for the CRs to autonomously and independently learn each of their optimal transmit power setting. As such, this work belongs to a class of notoriously challenging RL problems known as Weakly Acyclic Stochastic Dynamic Games for which there was no known RL algorithm with guaranteed convergence to the optimal policy. However, \cite{arslantcont2017} recently presented a modified general table-based Q-learning algorithm (a very common form of RL) for which convergence in an asymptotic infinite learning time was proved.

At the same time, the research area of RL have seen notable advances over the recent past years. The work \cite{mnih2015human} constitutes a major milestone by introducing deep Q-networks (DQNs), a new approach to Q-learning with better learning performance than conventional table-based Q-learning that is based on the use of deep neural networks to approximate the Q action-value function. This advance spurred research into the application of single-agent DQNs for a variety of applications, including wireless communications. In \cite{fang2017intelligent}, a single-agent power control Deep Q-learning (DQL) technique was introduced for an underlay CR system consisting of a single primary and a single secondary link. Also, the work \cite{xu2017deep} applied DQL to dynamic resource allocation in cloud Radio Access Networks, \cite{he2017deep} studied the application of DQL to control interference alignment, and \cite{sun2017learning} used DQL as a trainable function approximator for arbitrary resource allocation algorithms.

Our main contribution is to present a multi-agent DQL technique for distributed resource allocation in a CRN with no requirement for agents coordination. The key challenge addressed by our technique is that the presence of multiple active learning CR leads to a non-stationary environment due to the interaction of the learners through the shared wireless environment. Our technique is the first to present a multi-agent DQL technique for distributed CR resource allocation with proven convergence to the optimal policy in a non-stationary environment due to the uncoordinated interaction of the agents. The field of works researching multi-agent DQL for wireless communications is rapidly growing but still sparse and, in contrast to our proposed technique, have not focused on scenarios with a non-stationary environment because of allowing agent coordination or learning at a central node. In this regards, \cite{Meng2019} presents a power allocation technique for a cellular network using DQL based on training at a central node, and \cite{Nasirjsac19} also presents a DQN technique with centralized training based on the experiences gathered by all agents.

While our proposed DQL technique can achieve convergence to optimality when learning time tends to infinity (as is customary in RL), we will show through simulation results that in a finite learning time our technique reaches the optimal policy in nearly 70 \% of cases and yields a mean performance within 3 \% of the exhaustive search optimal solution. Moreover, as a second key contribution, we present a case that shows that the application of standard single-agent DQL in uncoordinated distributed CR resource allocation may not reach convergence (to any policy, optimal or not) due to the large noise present in Q-value estimation from the non-stationary environment.

\vspace*{-1mm}
\section{System Setup and Problem Formulation}\vspace*{-1mm}

We consider a system comprised of two networks that operate by simultaneously using the same radio spectrum band following an underlay DSA mechanism. In the system, a primary network (PN) is incumbent to the spectrum band in use and a secondary network (SN) is formed by CRs which are assumed to operate in a fully autonomous manner, i.e. there is no coordination between the CR nodes during resource allocation and no exchange of information between the two networks. By requirement of the underlay DSA operation, the CRs in the SN are limited in their transmit power so that the interference they create on the PN does not exceed an established limit. The primary network is comprised of $M$ access points (APs), each transmitting to one wireless receiver.



To operate following underlay DSA, each CR assesses the effect of its transmission on the PN, without any exchange of information between the PN and SN, by estimating the relative change in the throughput at its nearest PN link as in \cite{narxnn}. As such, owing to the one-to-one relation between Signal-to-Interference-plus-Noise ratio (SINR) and throughput, the CRs implement underlay DSA by avoiding to exceed a limit on the relative throughput change in their nearest PN link (understood hereafter to be the one that is received with largest power). Nodes in both wireless networks perform resource (power in this work) allocation in order to optimize their transmission. The increase in transmit power at a CR will increase the relative throughput change at its nearest PN link, risking to exceed the underlay DSA-imposed limit. Therefore, we seek an algorithm for each transmitting CR to find, without coordination with other CRs, its transmit power such that it achieves the largest possible throughput without exceeding the limit on relative throughput change on its nearest PN link.

\vspace*{-1mm}
\section{DQN-Based Distributed and Uncoordinated Multi-Agent Resource Allocation}\label{clcr}\vspace*{-1mm}

We adopt the use of RL in the form of Q-learning to solve the CR resource allocation problem described in the previous section because it enables model-free learning. The common principle to RL is that a learning agent is able to find a resource allocation policy by iteratively trying out each available action multiple times. Specifically, let $S = \{S_0,S_1,\dots,S_{n-1}\}$ be the set of environment states, and $A= \{a_1,a_2,\dots,a_m\}$ be the set of actions. At time $t$, the agent takes an action $a(t)\in A$ on the environment while it is in state $x_t\in S$. As a result, the agent receives an immediate reward $R(x,{a})$ that represents the effect of the selected action on the environment, and the system transitions to a new state $x_{t+1}\in S$. The same set of actions and states are assumed for all CRs. We define the state space, action space and reward function as follows:

\emph{\bf{Action space:}} Each CR will search over the discrete action space $A=\{0,p_0,p_1,\dots,p_L\}$ of possible transmit powers (where '0' indicates no transmission), for the best choice that maximizes throughput while meeting the underlay DSA constraint (limit on relative throughput change on a PN link). In our uncoordinated and distributed setup, each agent choose actions independently of the others.

\emph{\bf{State space: }} The state of the environment reflects whether the limits for underlay DSA are being met or not. As such, the environment is in state $S_0$ when the relative throughput change at all the PN links that are nearest to CRs transmissions are below the pre-decided limit, and is in state $S_1$ otherwise. It is assumed here that there is a control channel in the SN over which the CRs share a single bit of information indicating the state of their nearest PN link.

\emph{\bf{Reward: }}The reward function reflects each CR goal to maximize their throughput and is defined as a function of the state and the action taken to transition to a new state as,\vspace*{-2mm}
\begin{eqnarray}\label{eq_Reward}
R^{(i)}_t(a_t, x_t)\!=\!\left\{\begin{array} {ll}
10^{T_i}, & \textrm{if environment state is } S_0,\\[-.03in]
0, & \textrm{if environment state is } S_1,
\end{array} \right. \\ [-.13in] \nonumber
\end{eqnarray}
\noindent where $T_i$ is the throughput on the $i$th. SN link. This reward function was designed to augment the difference between rewards for actions that correspond to low SINRs. Also, we note that while the above is the reward function proper to our problem solution, in most of the simulation results we used as reward function $R=10^{\sum_i T_i}$, this is, the sum throughput in the SN. This is for fair comparison of our proposed solution against the optimal solution found through exhaustive search. To be optimal, the exhaustive search in the benchmark needs to be for a single global optimum, which is defined as the sum throughput in the SN.

With Q-learning, each agent finds the best policy $\pi^*$ that maximizes the expected sum of discounted rewards, \vspace*{-2mm}
\begin{eqnarray}
V(x,\pi)=\sum_{t=1}^{\infty} \gamma^{t}E[R_t|\pi,(x_0=x)], \\ [-.2in] \nonumber
\end{eqnarray}
where $x_0$ and $x$ denote the initial and the current state, $\gamma$ is the discount factor and $R_t$ is the reward for the agent at epoch $t$.  As in our case when the environment model is not immediately known, the expectation of the accumulated discounted rewards can be approximated through the Q-function using the time difference technique as, \vspace*{-2mm}
\begin{eqnarray}
Q_{t+1}(x_t, a_{t})=Q_{t}(x_t, a_{t})+\alpha_t\big[R_t+\gamma\max_{{a'\in A} } \nonumber\\[-.02in]
Q_t(x_{t+1},a')-Q_{t}(x_t, a_{t})\big], \label{qvalues1}\\[-.23in] \nonumber
\end{eqnarray}
where $\alpha_t \in (0,1)$ is the learning rate. In a table-based Q-learning algorithm, the Q-values $Q(s, a)$ are iteratively estimated using  \eqref{qvalues1} by exploring all actions multiple times from all states. However, this approach is impractical in that it involves a very large number of learning steps. The work in \cite{mnih2015human} constitutes a major milestone by introducing the deep Q-networks (DQNs), a new approach to Q-learning that learns the Q action-value function faster over large action and state spaces by using a deep neural network as an efficient nonlinear approximator. Also, DQN includes a technique known as ``experience replay'' to improve the learning performance, where at each time step, the tuple $e_i(t) = ({a_i}(t),x_i(t),r_i(t),x_i(t+1))$ is stored in a ``replay memory''. Moreover, in DQN each agent utilizes two separate neural networks: one as action-value function approximator $Q_i(x,{a};\theta_i)$ and another as target action-value function approximator $\hat{Q_i}(x,{a};\theta^-_i)$, where $\theta_i$ and $\theta^-_i$ denote the parameters (weights) for each neural network. At each learning step, the parameters $\theta_i$ of each agent's action-value function are updated through a mini-batch of random samples from the replay memory. The parameters $\theta_i$ are updated through a gradient descent backpropagation algorithm using the error function\vspace*{-2mm}
\begin{eqnarray}
\label{grad_dec}
\!L(\theta_i)\!\!=\!E[(r_i(x,a)\!+\!\gamma \underset{\hat{a}\in A}{\mathrm{\textstyle\max}} (\hat{Q}_i(\hat{x},\hat{a};\theta^-_i)) \!-\! Q_i(x,{a};\theta_i))^2].\!\! \\ [-.23in] \nonumber
\end{eqnarray}
Only every $c$ learning steps, the parameters $\theta^-_i$ are updated through the operation $\theta^-_i\! \leftarrow \!\theta_i$. This is to damp non-stationarity of the target values, which hampers convergence.

While the above DQN technique, which we will identify as ``standard DQL'', has shown significant success in many applications, it is a technique for single agent scenarios and as such, shares the positives of single agent Q-learning (e.g. guaranteed convergence to optimal policy as time tends to infinity) and its limitations for multi-agent scenarios. In our present case, the action of any agent may affect the environment of other agents (the transmission of one CR translates into interference to the rest of the SN and the PN), which compounded with a lack of agent coordination results in a multi-agent non-stationary environment for which standard DQL would not be adequate. Our problem of uncoordinated distributed multi-CR resource allocation belongs to a class of notoriously challenging RL problems known as Weakly Acyclic Stochastic Dynamic Games for which there was no known RL algorithm with guaranteed convergence to the optimal policy. However, \cite{arslantcont2017} recently presented a general table-based Q-learning algorithm for which it was proved convergence with probability one in an asymptotically infinite learning time. We use this work as a basis to present next a novel DQL configuration that shares the positive qualities from standard DQL and inherits the convergence guarantee from the table-based Q-learning algorithm in \cite{arslantcont2017}.

The main obstacle to convergence in the problem at hand is the presence of multiple active learners interacting through their entanglement over the environment, yielding a non-stationary environment. Standard DQL already incorporates a mechanism to address non-stationarity of the target action-value function during learning. However, this is not enough for the problem at hand because the non-stationary environment results in ``noisy'' and non-stationary observations of the reward which, as can be seen in \eqref{grad_dec}, leads to non-convergence learning of the DQL neural network. To address the non-stationary environment issue in table-based Q-learning, \cite{arslantcont2017} introduced the idea that learning occurs over a succession of multiple ``exploration phases'' where in each of them agents take the action determined by the current policy most of the time and only occasionally (as determined by a random draw with an ``experimentation probability'') it explores a different action. This results in agents experiencing near-stationary environments (with stationarity occasionally broken by exploration of non-current policy actions by other agents), which allows the agents to accurately estimate their Q-values. In addition, it is necessary to also implement a mechanism for gradual learning of the best policy that does not require coordination between agents. In \cite{arslantcont2017}, the technique proposed for this purpose is a mechanism called ``Best Reply Process with Inertia''. In this mechanism, after each exploration phase a policy is selected from a candidate set formed from those actions associated with a Q-value that is within a tolerance range of the largest Q-value. Inertia in the choice is implemented by keeping the current action if it belongs to the candidate set or, otherwise, keeping the current action with an inertia probability $\lambda$ (usually larger than 0.5), and choosing one policy from the candidate set with probability $1-\lambda$.

Algorithm \ref{alg1} shows all these ideas together. In the algorithm, lines 9 to 14 comprise an exploration phase. Our implementation of inertia (lines 15 and 16) differs from  \cite{arslantcont2017} by always choosing a policy from the random draw with probability $\lambda$. The tolerance level $\delta^i$ used in line 15 is also particular to our solution and is calculated as three times the largest moving standard deviation among Q-values. Within the exploration phase, line 10 shows the choice of action and lines 11, 12, and 13 comprise the DQN. We eliminated the replay memory because it operates counter to the intend of an exploration phase and we eliminated the target action-value neural network, replacing it by an array that stores Q-values.

\vspace*{-0.1mm}
\begin{algorithm}
 	\caption{Uncoordinated Distributed Multi-agent DQN}\vspace*{-.3mm} \label{alg1}
 	\begin{algorithmic}[1]
 	    \STATE Set parameters \vspace*{-.4mm}
 	    \STATE $\rho\in$ {(0,1)} : {Experimentation probability} \vspace*{-.4mm}
 	    \STATE $\lambda\in$ {(0,1)} : {Inertia} \vspace*{-.4mm}
 	    \STATE $\gamma\in$ {(0,1)} : {Discount factor}
 	    \STATE Initialize policy $\pi_0 \in \Pi$ {(arbitrary)} \vspace*{-.4mm}
        \STATE Sense state $x_0$ \vspace*{-.4mm}
 		\STATE 	Initialization of the neural network for action-value function $Q_i$ with random weights $\theta_i$ \vspace*{-.4mm}
 		
 		\FOR { $0\leq k \leq K$}
 	    \FOR {Iterate $t= t_k, t_k +1,\cdots, t_{k+1} - 1$}
 	    \STATE ($k$th. exploration phase)\\
 	            $ a_{t} =
        \begin{cases}
       \text{$\pi_k(x_t), \hspace{2mm} w.p.\hspace{2mm} 1-\rho $ }\\[-.05in]
       \text{$ \textrm{any } a \in {A},\hspace{2mm} w.p.\hspace{2mm} \rho/|A| $}
       \end{cases} $

 		\STATE Update the state $x^{(i)}_{t+1}$  and the reward $R^{(i)}_t$.
 		
 		\STATE Update parameters ($\theta$) of action-value function $Q(x_t^{(i)}, a_t^{(i)};\theta_i)$, through mini-batch backpropagation with error function \eqref{grad_dec} \vspace*{-.4mm}
 		\STATE every $c$ step update array in memory with target action-value function: $\hat{Q}(x,a) \leftarrow Q(x,{a};\theta_i)$, $\forall x, a$. \vspace*{-.4mm}

        \ENDFOR \vspace*{-.4mm}
 	
        \STATE
        $\Pi_{k+1}^i = \{\hat{\pi}^i \in\Pi^i:Q_{t_{k+1}}^i(x,\hat{\pi}^i(x))$\\
        $\qquad \qquad \ge\max_{v^i}Q_{t_{k+1}}^i (x, v^i)-\delta^i, \text {for all x}\}$

        \STATE $ \pi_{k+1}^i =
        \begin{cases}
       \text{$\pi_k^i, \hspace{4mm} w.p.\hspace{4mm} \lambda $ }\\[-.05in]
       \text{$ \mbox{any } \pi^i \in \Pi_{k+1}^i, w.p. \frac {(1-\lambda)} {|\Pi_{k+1}^i|}$}
       \end{cases} $
        \ENDFOR \vspace*{-.4mm}
 	\end{algorithmic}
 \end{algorithm}
\vspace*{-4mm}

\emph{Remark:} Because our multi-agent DQN maintains all the elements of the table-based Q-learning algorithm in \cite{arslantcont2017} needed for convergence, we state that our DQL technique achieves convergence to optimality with probability one as learning time tends to infinity, with proof as shown in \cite{arslantcont2017}.

\vspace*{-1mm}
\section{Experimental Results and Discussion}\label{simulation}\vspace*{-1mm}


We evaluated the proposed multi-agent DQL technique through 100 Monte Carlo simulations runs where a PN shares through underlay DSA a 180 kHz radio spectrum band with an SN formed by CRs implementing the proposed technique. The PN consists of nine access points (APs) organized in a three-by-three grid (wrapping around all edges to avoid edge effects), with each AP separated by a distance of 200 m. Of the nine APs, only seven (chosen at random for each Monte Carlo run) are active, each transmitting to one receiver located at random within the coverage area of the corresponding AP. The SN consists of two links with each of the two CR transmitters located at random anywhere within the PN three-by-three grid and the two receivers randomly located within 50 m of their corresponding transmitters. This setup is intended to broadly model a network of small radio devices (the SN) sharing spectrum with an incumbent network (the PN) of larger devices. For underlay spectrum sharing, the limit on relative throughput change on the PN links was set to 5 \%.

In both networks, the received signal power $P_{Rx}$ undergoes path, penetration and shadowing loss, modeled for a link length of $d$ km as $P_{Rx} = P_{Tx} - 128.1 -37.6 \log d -10 -S$, \cite{3GPP_TR_36_814}, where all magnitudes are in dBs, $P_{Tx}$ is the transmit power, $S$ is the shadowing loss (modeled as a zero-mean Gaussian random variable with 6 dB standard deviation) and the penetration loss is fixed at 10 dB. Also, all channels have AWGN power -130 dBm and they present a delay spread that follows the Pedestrian B typical urban model from \cite{ITUchmod}. In addition, all transmissions in the PN and SN implement adaptive transmit power and adaptive modulation and coding (AMC). We assumed that PN transmitters can adapt transmit power in the range of -30 and +20 dBm following the iterative power allocation algorithm in \cite{Chawla1999} and the CRs operate with an action space was defined as the fourteen transmit power levels between -10 and 20 dBm equally spaced by 2.5 dBm, plus no transmission (transmit power zero). For simulation purposes, we adopt the AMC scheme defined for LTE (fifteen possible modes combining different channel coding rates and one of QPSK, 16-QAM or 64-QAM modulation scheme), \cite{3GPP_TR_36_213}.

Following evaluation of multiple configurations we implemented the DQL neural network as a five-layer multilayer perceptron (MLP) with input being the environment state and outputs the Q-values for each possible action. Then, the MLP output layer was formed by fourteen neurons with a linear activation function. Also, the implemented MLP had three, five and seven neurons in the hidden layers (from input to output). The activation function for all hidden layers was a saturated ReLU (rectified linear unit). Training for the DQL neural network was done through backpropagation with the gradient descent algorithm configured (unless otherwise noted in specific experiments) with learning rate equal to 0.01, mini-batch size of 60 training samples consisting of tuples $<$\emph{current state, next state, action taken, target Q-value}$>$, and epoch (equivalent to one exploration phase) length equal to 60. The neural network weights are initialized with random values uniformly distributed between 0 and 1. 

Unless otherwise noted for specific experiments, the proposed DQL technique used the following default parameters: number of exploration phases equal to 60, experimentation probability $\rho=0.15$, inertia probability $\lambda=0.35$, update target Q-values every $c=30$ training steps.

\begin{figure}[tbhp]
	\centering
	\includegraphics[width=0.37\textwidth]{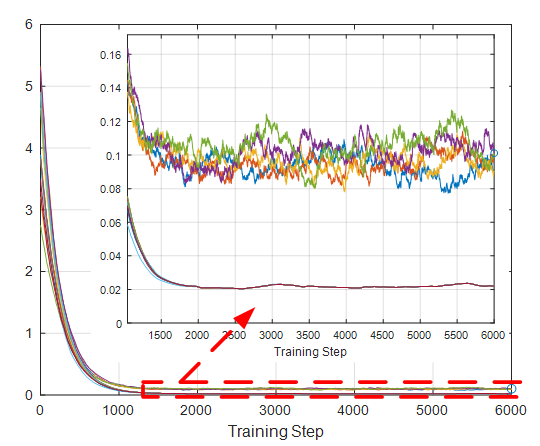}\vspace*{-1mm}
	\caption{Q-values evolution during multi-agent learning with standard DQL.}
	\label{singleagent}\vspace*{-1mm}
\end{figure}

\begin{figure}[tbhp]
	\centering
	\includegraphics[width=0.37\textwidth]{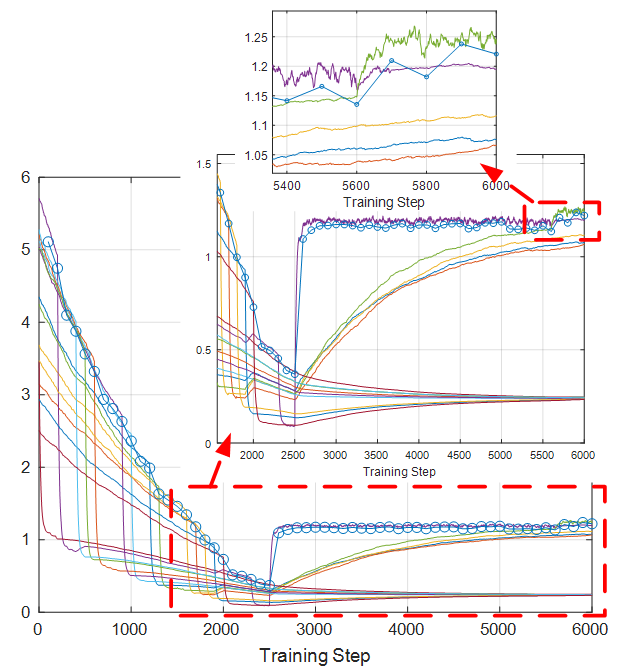}\vspace*{-1mm}
	\caption{For identical settings as in Fig. \ref{singleagent}, Q-values evolution using the proposed DQL approach (solution converges to optimal policy).}
	\label{multiagent}\vspace*{-1mm}
\end{figure}

Figs. \ref{singleagent} and \ref{multiagent} compare the RL evolution for, respectively, standard DQL and our proposed approach for the same exemplar system realization. Each curve in both figures shows for CR link 2 the evolution over the training steps of the Q-value for one action when in state $S_0$, the exception  being the curve with circular markers, which shows the tolerance level $\delta^i$ (the action associated with a Q-value larger than this curve will be a potential next-policy). Inserts in the figures show zoomed-in portions of the curves. For standard DQL, Fig. \ref{singleagent} shows how the non-stationary environment results in ``noisy'' Q-values which leads to a non-convergent behavior. These noisy Q-values are because the action exploration of the two CR transmitters results in different pairings of actions from each agent and, consequently, different rewards (and possibly next state) for the same initial state and action. In the particular case of Fig. \ref{singleagent}, the other CR frequently tests actions that transition the system from state $S_0$ to state $S_1$, resulting in all Q-values in CR link 2 tending to a value near zero. However, the Q-values do not converge to zero and become noisy because a few of the actions being randomly explored keep the system in state $S_0$ and yield a reward larger than zero. The end result is that the algorithm is unable to converge and the final policy becomes practically a random choice between four different actions. Importantly, the configuration depicted in Fig. \ref{singleagent} makes use of the customary separate target action-value set updated every $c=60$ training steps, which on the surface could have been considered to provide some level of robustness against the non-stationary environment. However, the important conclusion from Fig. \ref{singleagent} is that standard DQL may not achieve convergence an uncoordinated multi-agent scenario. In contrast, Fig. \ref{multiagent} depicts the learning evolution under completely identical settings for our proposed multi-agent DQL technique and illustrates success in achieving convergence to the optimal solution. As can be seen in the zoomed-in details of the curve (two zoom levels are shown in the figure), the near-stationary environment achieved during exploration phases significantly reduces the ``noise'' in Q-values (the Q-value corresponding to the active policy presents more noise than the others because it is explored more frequently), yielding low-noise and discernible values. Fig. \ref{multiagent} also illustrates how convergence to the optimal policy may occur near the end of the simulation and provides a sense that for our finite time simulations there may be cases where convergence to the optimal policy may not have occurred yet. However, as will be seen, we choose parameters so that the performance effect in these occurrences is minimal.

\vspace*{-5mm}
\vspace*{-1mm}

\vspace*{-4mm}
\vspace*{-1mm}

Table \ref{ResultsTab} shows the simulation results where we evaluated the ability of the proposed technique to find the multi-agent resource allocation optimal solution in a finite learning time. In the experiments, the optimal solution was calculated by centralized exhaustive search over all transmit power combinations for both CR links that yields the largest sum throughput in the SN while meeting the underlay DSA limit in the PN. While the proposed DQL technique will arrive at the optimal solution, this result is guaranteed to happen for the generality of cases when the learning time is allowed to tend to infinity. However, practical deployment in the field and simulation limitations, dictates a finite limit to the learning time. Consequently, we measured the percentage of cases that the proposed algorithm found the optimal policy and also the mean relative absolute difference in the SN sum throughput between the optimal solution and our proposed technique. In addition, we measured the mean number of exploration phases needed to reach convergence (counting the cases that do not achieve convergence in the given finite time with the maximum number of exploration phases). The default parameter settings in our DQL technique detailed earlier were chosen to provide very small mean relative difference within a reasonable span of time and its results are shown in Table \ref{ResultsTab} as ``Default setting''.

\vspace{-4mm}
\begin{table}[tbh]
	\centering
	\begin{tabular} { |m{12.5em} | c | c | c |  }
		\hline 
		\vspace{-4mm} \thead{Description} & \thead{Mean \\ Relative \\ Difference} & \thead{Mean \\ Exp. Phases \\ to Converge} & \thead{Percent \\ Optimal \\ Policy} \\  [-0.6ex]
		\hline		
		Default setting & 0.0249 & 33.62 & 69 \\
		\hline
		Standard DQL $c=1$  & 0.0945 & N/A & 56  \\
		\hline
		Standard DQL $c=60$  & 0.0744 & N/A & 55  \\
		\hline
		$c=1$  & 0.0331 & 35.18 & 66  \\
		\hline
		$c=60$  & 0.0096 & 33.42 & 72  \\
		\hline
		Mini batch size = 120  & 0.0375 & 35.6 & 67  \\
		\hline
		Mini batch size = 30 & 0.0241 & 36.35 &  69 \\
		\hline
		Uncoordinated per-CR reward & 0.0411  & 32.01 &  70  \\
		\hline
	\end{tabular}\vspace{1mm}
	\caption{Results of Experiments} \label{ResultsTab}
\end{table}
\vspace{-4mm}

The experiments described in Table \ref{ResultsTab} ``Standard DQL $c=1$'' and ``Standard DQL $c=60$'' present results for the case of direct application of a standard (single agent) DQL technique with different settings for the parameter $c$ that measures the number of training steps in-between refreshing the target Q-values. The results show that in close to half of the cases this algorithm does not converge to the optimal policy. The much larger mean relative difference compared with our proposed technique indicates that many cases will never convergence (as was exemplified in Fig. \ref{singleagent}). These results shows the inadequacy of standard DQL for the considered multi-agent scenario and Validates our proposed DQL approach.

The rest of experimental results shown in Table \ref{ResultsTab} study the performance of our proposed DQL scheme with different parameters. The ``$c=1$'' and ``$c=60$'' cases show that a larger value for the number of steps between refreshing the target Q-values provides some performance improvement. The results described as ``Mini batch size = 120'' and ``Mini batch size = 30'' show little performance difference with the mini batch size used in the default setting. In order to compare viz-a-viz to the optimal exhaustive search solutions, all the results presented so far correspond to the use of a reward function that is a function of the sum throughput across all CRs. However, the actual implementation of the uncoordinated DQL would not have the CR sharing their throughput to calculate their sum, and instead, would have each CR calculate a reward as in \eqref{eq_Reward} based on their individual throughput. The results for this case are labeled ``Uncoordinated per-CR reward''. As can be seen, performance is similar to that of our default setting (with SN sum throughput-based reward), confirming the success of our proposed DQL technique, although the mean relative difference is slightly larger in the per-CR reward. This is because of occasional cases where each CR attempts to maximize their own throughput by increasing transmit power with such a coupling between the CRs that the interference on each other results in a sum throughput slightly lower than the optimal. However, we emphasize that the use of sum throughput as a metric was dictated by the need to define a comparative benchmark with a single global optimum, and is not a performance measure that fully corresponds to uncoordinated conditions.

\vspace*{-2mm}
\section{Conclusion}\label{conclsec}\vspace*{-1mm}

In this paper, we presented a novel deep reinforcement learning technique capable for the first time of achieving convergence to the optimal solution in the case of uncoordinated interacting multiple-agent CRs. The presented novel DQL technique succeeds in addressing the challenge of a non-stationary multi-agent environment that results from the dynamic interaction between radios through the shared wireless environment in underlay DSA. Simulation results show that under a finite learning time the presented technique finds the optimal policy in nearly 70 \% of cases and yields performance within 3\% of an exhaustive search optimal solution. We also present a case that shows that standard single-agent deep reinforcement learning may not achieve convergence when used in a non-coordinated, coupled multi-radio scenario.

\vspace*{-1mm}
%
\bibliographystyle{IEEEtran}
\bibliography{spawc_2020}

\begin{thebibliography}{10}
\providecommand{\url}[1]{#1}
\csname url@samestyle\endcsname
\providecommand{\newblock}{\relax}
\providecommand{\bibinfo}[2]{#2}
\providecommand{\BIBentrySTDinterwordspacing}{\spaceskip=0pt\relax}
\providecommand{\BIBentryALTinterwordstretchfactor}{4}
\providecommand{\BIBentryALTinterwordspacing}{\spaceskip=\fontdimen2\font plus
\BIBentryALTinterwordstretchfactor\fontdimen3\font minus
  \fontdimen4\font\relax}
\providecommand{\BIBforeignlanguage}[2]{{%
\expandafter\ifx\csname l@#1\endcsname\relax
\typeout{** WARNING: IEEEtran.bst: No hyphenation pattern has been}%
\typeout{** loaded for the language `#1'. Using the pattern for}%
\typeout{** the default language instead.}%
\else
\language=\csname l@#1\endcsname
\fi
#2}}
\providecommand{\BIBdecl}{\relax}
\BIBdecl

\bibitem{arslantcont2017}
G.~Arslan and S.~Y{\"u}ksel, ``Decentralized q-learning for stochastic teams
  and games,'' \emph{IEEE Transactions on Automatic Control}, vol.~62, no.~4,
  pp. 1545--1558, April 2017.

\bibitem{mnih2015human}
V.~Mnih, K.~Kavukcuoglu, D.~Silver, A.~A. Rusu \emph{et~al.}, ``Human-level
  control through deep reinforcement learning,'' \emph{Nature}, vol. 518, no.
  7540, p. 529, 2015.

\bibitem{fang2017intelligent}
X.~{Li}, J.~{Fang}, W.~{Cheng}, H.~{Duan} \emph{et~al.}, ``Intelligent power
  control for spectrum sharing in cognitive radios: A deep reinforcement
  learning approach,'' \emph{IEEE Access}, vol.~6, pp. 25\,463--25\,473, 2018.

\bibitem{xu2017deep}
Z.~Xu, Y.~Wang, J.~Tang, J.~Wang, and M.~C. Gursoy, ``A deep reinforcement
  learning based framework for power-efficient resource allocation in cloud
  rans,'' in \emph{Communications (ICC), 2017 IEEE International Conference
  on}.\hskip 1em plus 0.5em minus 0.4em\relax IEEE, 2017, pp. 1--6.

\bibitem{he2017deep}
Y.~He, Z.~Zhang, F.~R. Yu, N.~Zhao \emph{et~al.}, ``Deep reinforcement
  learning-based optimization for cache-enabled opportunistic interference
  alignment wireless networks,'' \emph{IEEE Transactions on Vehicular
  Technology}, vol.~66, no.~11, pp. 10\,433--10\,445, 2017.

\bibitem{sun2017learning}
H.~Sun, X.~Chen, Q.~Shi, M.~Hong, X.~Fu, and N.~D. Sidiropoulos, ``Learning to
  optimize: Training deep neural networks for wireless resource management,''
  in \emph{IEEE 18th Int. Workshop on Signal Proc. Adv. in Wireless Comm.
  (SPAWC)}, 2017, pp. 1--6.

\bibitem{Meng2019}
F.~{Meng}, P.~{Chen}, and L.~{Wu}, ``Power allocation in multi-user cellular
  networks with deep q learning approach,'' in \emph{IEEE International
  Conference on Communications (ICC)}, May 2019, pp. 1--6.

\bibitem{Nasirjsac19}
Y.~S. {Nasir} and D.~{Guo}, ``Multi-agent deep reinforcement learning for
  dynamic power allocation in wireless networks,'' \emph{IEEE J. on Sel. Areas
  in Comm.}, vol.~37, no.~10, pp. 2239--2250, Oct 2019.

\bibitem{narxnn}
\BIBentryALTinterwordspacing
F.~S. Mohammadi and A.~Kwasinski, ``Neural network cognitive engine for
  autonomous and distributed underlay dynamic spectrum access,'' \emph{CoRR},
  vol. abs/1806.11038, 2018. [Online]. Available:
  \url{http://arxiv.org/abs/1806.11038}
\BIBentrySTDinterwordspacing

\bibitem{3GPP_TR_36_814}
``Further advancements for e-utra physical layer aspects, document 3gpp tr
  36.814 v9. 0.0,'' \emph{3rd Generation Partnership Project; Technical
  Specification Group Radio Access Network; E-UTRA}, 2010.

\bibitem{ITUchmod}
``Guidelines for evaluation of radio transmission technologies for imt-2000,''
  \emph{Rec. ITU-R M. 1225}, 1997.

\bibitem{Chawla1999}
{Xiaoxin Qiu} and K.~{Chawla}, ``On the performance of adaptive modulation in
  cellular systems,'' \emph{IEEE Transactions on Communications}, vol.~47,
  no.~6, pp. 884--895, June 1999.

\bibitem{3GPP_TR_36_213}
``Physical layer procedures, document 3gpp tr 36.213 v9. 2.0,'' \emph{3rd
  Generation Partnership Project; Technical Specification Group Radio Access
  Network; E-UTRA}, 2010.

\end{thebibliography}

\end{document}